\documentclass[onecollarge,natbib]{svjour2}
\bibpunct{[}{]}{;}{n}{}{,} 
\smartqed  
\usepackage{graphicx}
\usepackage{color}

\renewcommand{\vec}[1]{\mbox{\boldmath $#1$}}
\journalname{Few-Body Systems}
\begin{document}

\title{Are there good probes for the 
di-neutron correlation in light neutron-rich nuclei?} 



\author{K. Hagino         \and
        H. Sagawa 
}


\institute{
K. Hagino \at
Department of Physics, Tohoku University, Sendai 980-8578, Japan \\
\and
K. Hagino \at
Research Center for Electron Photon Science, Tohoku University, 
Sendai 982-0826, Japan \\
\and
K. Hagino \at
National Astronomical Observatory of Japan, 2-21-1 Osawa, Mitaka, 
Tokyo 181-8588, Japan \\
\and
H. Sagawa\at
RIKEN Nishina Center, Wako 351-0198, Japan \\
\and
H. Sagawa\at
Center for Mathematics and Physics, University of Aizu, Aizu-Wakamatsu, 
Fukushima 965-8560, Japan
}

\date{Received: date / Accepted: date}

\maketitle

\begin{abstract}
The di-neutron correlation is a spatial correlation with which two 
valence neutrons are located at a similar position inside a nucleus. 
We discuss possible experimental probes for the di-neutron correlation. 
This includes the Coulomb breakup and the pair transfer 
reactions of neutron-rich nuclei, 
and the direct two-neutron decays of nuclei beyond the neutron drip-line. 
\keywords{Neutron-rich nuclei\and Di-neutron correlation \and Coulomb breakup 
\and pair transfer reactions \and two-neutron decays}
\end{abstract}

\section{Introduction}

One of the most important issues in many-body physics is to clarify 
the nature of correlations beyond the independent particle picture. 
In nuclear physics, 
the pairing correlation has been well recognized as 
a typical many-boy correlation\cite{RS80,BB05}, 
which 
leads to a characteristic even-odd staggering of biding energy providing 
an extra binding for even-mass nuclei. 
A pairing interaction scatters 
nucleon pairs from a single-particle level below the Fermi surface to 
those above, and as a consequence, each single-particle level
is occupied with a fractional occupation probability. 

With the pairing correlation, one may naively 
expect that two nucleons 
forming a pair are located at a similar position inside a nucleus. 
A spatial structure of two valence neutrons 
has in fact attracted much attention in the past. 
One of the earliest publications on this problem is by 
Bertsch, Broglia, and Riedel, who solved a shell model for $^{210}$Pb 
and showed that the two valence neutrons 
are strongly clusterized \cite{BBR67}. 
Subsequently, Migdal argued that two neutrons may 
be bound in a nucleus even though 
they are not bound in the vacuum \cite{M73}. 

The strong localization of two neutrons inside a nucleus has been 
referred to as the {\it di-neutron correlation}. 
It has been nicely demonstrated in 
Ref. \cite{CIMV84} that 
an admixture of confiurations of single-particle orbits with 
opposite parity is essential to create the strong  di-neutron 
correlation.
This implies that the pairing correlation acting only on 
single-particle 
orbits with the same parity is not sufficient in order 
to develop the di-neutron correlation, 
and the pairing model space needs to be taken sufficiently largely so that 
both positive parity and negative parity states are included. 

\begin{figure*}
\centering
  \includegraphics[width=0.4\textwidth]{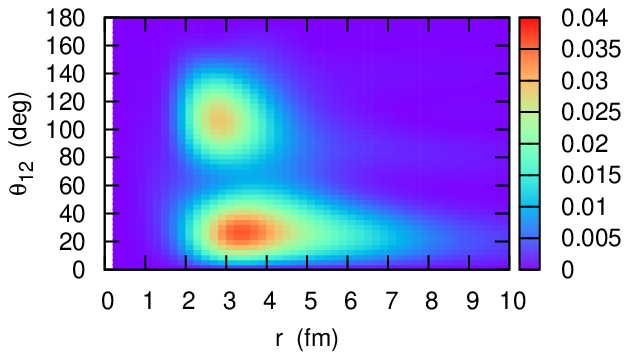}
  \includegraphics[width=0.4\textwidth]{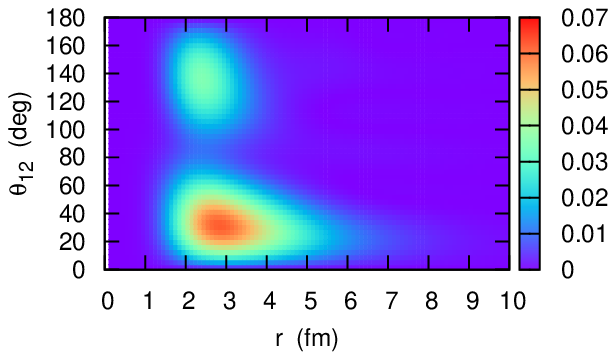}
\caption{
The two-particle densities for $^{11}$Li (the left panel) and for $^6$He (the 
right panel) obtained with a three-body model calculation 
with a density dependent contact pairing interaction \cite{HS05}. 
These are plotted as a function of neutron-core distance, $r_1=r_2\equiv r$, 
and the opening angle between the valence neutrons, $\theta_{12}$. 
The densities are weighted with a factor $8\pi^2r^4\sin\theta_{12}$. } 
\label{density}    
\end{figure*}

Although the di-neutron correlation exists even in stable nuclei, 
it is therefore more enhanced in weakly bound nuclei 
because the admixtures of single-particle orbits with different 
parities are easier due to the couplings to the continuum 
spectra\cite{HTS13,SH15}. 
Three-body model calculations have revealed that 
a strong di-neutron correlation
indeed exists in weakly-bound 
Borromean nuclei, such as $^{11}$Li and $^6$He 
\cite{BE91,Zhukov93,BBBCV01,HS05,HSCS07,KKM10}. 
For instance, Fig. \ref{density} shows the two-particle density for 
the $^{11}$Li and $^6$He nuclei obtained with the three-body model 
calculation with a density dependent contact pairing interaction \cite{HS05}. 
One can see that the densities are concentrated in the region with 
small opening angles, 
that is nothing but the di-neutron correlation. 
It has been shown that the di-neutron correlation exists also in heavier 
neutron-rich nuclei \cite{MMS05,PSS07}
as well as in infinite neutron matter \cite{M06}.
The di-proton correlation, which is a counter part of the di-neutron 
correlation, has also been shown to 
exist in the proton-rich 
Borromean nucleus, $^{17}$Ne\cite{OHS10}. 

From these studies, the di-neutron correlation seems to have 
been theoretically established. However, it is not straightforward 
to probe it experimentally. In this contribution, we discuss 
how one can probe the di-neutron correlation. To be more specific, 
we shall discuss 
the Coulomb breakup, 
the two-neutron transfer reactions, and 
the two-nucleon 
radioactivity as possible probes for the correlation. 

\section{Coulomb breakup of Borromean nuclei}

Let us first discuss the Coulomb breakup reactions of Borromean nuclei, 
$^{11}$Li and $^6$He, 
for which the experimental data have been available in Refs. \cite{N06,A99}. 
Those experimental breakup cross sections, 
especially those for the $^{11}$Li nucleus, show 
a strong concentration in the low excitation region, 
reflecting the halo structure of these nuclei. 
Moreover, the experimental data for $^{11}$Li are consistent 
only with the 
theoretical calculation 
which takes into account 
the interaction between the valence 
neutrons, strongly 
suggesting the existence of the di-neutron correlation in 
this nucleus (see also Ref. \cite{EHMS07}). 

For the Coulomb breakup of Borromean nuclei, one can go one step 
further, given that 
the Coulomb breakup process takes place predominantly 
by the dipole excitation. 
The Coulomb breakup cross sections 
with the absorption of dipole photons are given by 
\begin{equation}
\frac{d\sigma_\gamma}{dE_\gamma}
=\frac{16\pi^3}{9\hbar c}
N_\gamma(E_\gamma)\cdot\frac{dB({\rm E1})}{dE_\gamma},
\label{sigma_gamma}
\end{equation}
where $N_\gamma$ is the number of virtual 
photons, and 
\begin{equation}
\frac{dB({\rm E1})}{dE_\gamma}
=
\frac{1}{2I_i+1}
\left|\langle \psi_f||D||\psi_i\rangle\right|^2
\delta(E_f-E_i- E_\gamma),
\label{BE1}
\end{equation}
is the reduced $E1$ transition probability.  
In this equation, $\psi_i$ and $\psi_f$ are the wave functions for 
the initial and the final states, respectively, 
$I_i$ is the spin of the initial state, and 
$D_\mu$ is the 
operator for the $E1$ transition. 
For the Borromean nuclei, assuming a three-body structure with 
an inert core, the $E1$ operator $D_\mu$ reads \cite{EB92},
\begin{equation}
\hat{D}_\mu = \frac{e_{\rm E1}}{2}\,(r_1Y_{1\mu}(\hat{\vec{r}}_1)+r_2Y_{1\mu}(\hat{\vec{r}}_2)), 
\label{E1}
\end{equation}
where 
 the 
E1 effective charge is given by
\begin{equation}
e_{\rm E1}=\frac{2Z_c}{A_c+2}\,e,
\end{equation}
with $A_c$ and $Z_c$ being the mass and charge numbers for the core 
nucleus. $\vec{r}_1$ and $\vec{r}_2$ are the coordinates 
for the valence neutrons. 

\begin{figure}
\centering
  \includegraphics[width=0.3\textwidth]{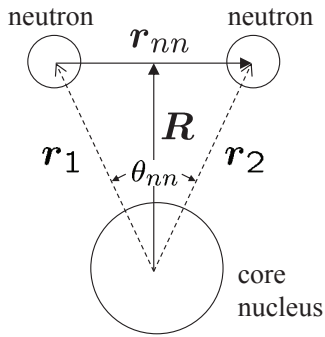}
\caption{
The geometry of a 2n halo nucleus consisting of a core nucleus and 
two valence neutrons.  }
\label{geometry}    
\end{figure}

Using Eq. (\ref{BE1}) and the closure relation for the final state, 
it is easy to derive that the total E1 strength 
(that is, the non-energy weighted sum rule) is proportional to the expectation 
value of 
the center of mass coordinate for the two valence neutrons, 
$R^2$, with respect to the ground state, that is, 
\begin{equation}
B({\rm E1})=\int dE_\gamma \frac{dB({\rm E1})}{dE_\gamma}
=\sum_f\frac{1}{2I_i+1}
\left|\langle \psi_f||D||\psi_i\rangle\right|^2
=\frac{3}{4\pi}\,e_{\rm E1}^2\langle \vec{R}^2\rangle, 
\end{equation}
with
$\vec{R}=(\vec{r}_1+\vec{r}_2)/2$.  
Even though the $B({\rm E1})$ strength distribution inevitably reflects 
both the correlation in the ground state and that in the final state, 
it is remarkable that one can extract the information which reflects 
{\it solely} the ground state properties after summing all the strength 
distribution. 
This implies that 
the average value of 
the opening angle between the valence neutrons can be directly 
extracted from the measured total $B({\rm E1})$ value once the 
root-mean-square distance between the valence neutrons, 
$\langle r^2_{nn}\rangle$, is available (see Fig. \ref{geometry}). 

This quantity is related 
to the matter radius and $\langle R^2\rangle$ 
in the  three-body model as \cite{BE91,EHMS07,VinhMau96}, 
\begin{equation}
\langle r_m^2\rangle = 
\frac{A_c}{A}\,\langle r_m^2\rangle_{A_c}
+\frac{2A_c}{A^2}\,\langle R^2\rangle 
+\frac{1}{2A}\,\langle r_{nn}^2\rangle, 
\label{eq:rnn}
\end{equation}
where $A=A_c+2$ is the mass number of the whole nucleus. 
The matter radii 
$\langle r_m^2\rangle$ can be estimated from interaction cross
sections. Employing the Glauber theory in the optical limit, Tanihata
{\it et al.} have obtained 
$\sqrt{\langle r_m^2\rangle}$ = 1.57 $\pm$ 0.04, 
2.48 $\pm$ 0.03, 2.32 $\pm$ 0.02, and 
3.12 $\pm$ 0.16 fm for $^4$He, $^6$He, $^9$Li, and 
$^{11}$Li, respectively \cite{Tani85,Ozawa01}.
Using these values, we obtain the 
rms neutron-neutron distance of 
$\sqrt{\langle r_{nn}^2\rangle}$ = 3.75 $\pm$ 0.93 and 
5.50 $\pm$ 2.24 fm for $^6$He and $^{11}$Li, respectively. 
Combining these values with the rms core - di-neutron distance, 
$\sqrt{\langle R^2\rangle}$, 
we obtain the mean opening angle of $\langle\theta_{nn}\rangle$ =
51.56$^{+11.2}_{-12.4}$ and 56.2$^{+17.8}_{-21.3}$ degrees for 
$^6$He and $^{11}$Li, respectively \cite{HS07}. 
These values are comparable to 
the result of the three-body model calculation, 
$\langle\theta_{nn}\rangle$=66.33 and 65.29 degree for $^6$He and 
$^{11}$Li, respectively \cite{HS05}, although 
the experimental values are somewhat smaller.

An alternative way to 
 extract the value  for $\sqrt{\langle r^2_{nn}\rangle}$ has been
 proposed which uses 
the three-body 
correlation study in the dissociation of two neutrons in halo nuclei 
\cite{M00}.  The two neutron correlation function provides the experimental
values for $\sqrt{\langle r_{nn}^2\rangle}$ to be  5.9 $\pm$ 1.2 and 6.6 $\pm$ 1.5
fm for $^6$He, $^{11}$Li, respectively \cite{M00}.
Bertulani and Hussein used these values
to estimate the mean opening
angles and 
obtained 
$\langle\theta_{nn}\rangle$=83 $^{+20}_{-10}$ and 
66 $^{+22}_{-18}$ degrees 
for $^6$He and $^{11}$Li, respectively \cite{BH07}. 
After correcting the effect of Pauli forbidden transitions using the 
method presented in Ref. \cite{EHMS07}, 
these values are slightly altered to be 
$\langle\theta_{nn}\rangle$=74.5 $^{+11.2}_{-13.1}$ and 
65.2 $^{+11.4}_{-13.0}$ degrees 
for $^6$He and $^{11}$Li, respectively \cite{HS07}. 
Notice that these values are
in a better agreement with the results of the three-body calculation
\cite{HS05}, especially for the $^6$He nucleus. 

In the absence of the correlations, the mean opening angle is 
exactly 
$\langle\theta_{nn}\rangle$=90 degrees. The extracted values of 
$\langle\theta_{nn}\rangle$ are significantly smaller than this value both 
for $^{11}$Li and $^6$He, providing a direct proof of the existence 
of the di-neutron correlation in these nuclei. A small drawback is that 
this method provides only the average value of 
$\langle\theta_{nn}\rangle$ and a detailed distribution is inaccessible. 
In reality, the mean opening angle is most probably 
an average of a smaller and
a larger correlation angles in the density distribution, as has been
shown in Fig. \ref{density}. 

\section{Two-neutron transfer reactions}

It has been recognized for a long time 
that two-neutron transfer reactions 
are sensitive to the pairing correlation\cite{Y62,OV01,Potel13}. 
The probability for the two-neutron transfer process is enhanced 
as compared to a naive expectation of sequential transfer process, 
that is, the square of one-neutron transfer 
probability\cite{CSP11,Montanari14}. 
The enhancement of pair transfer probability has been attributed to 
the pairing effect, such as 
the surface localization of a Cooper 
pair\cite{CIMV84,ILM89}. 
The pair transfer reaction is thus considered to provide 
a promising way to probe
the di-neutron correlation. 
However, the reaction dynamics is rather complicated 
and has not yet been well established. 
For instance, it is only with a recent calculation 
that a theoretical calculation achieves a satisfactory agreement 
with the experimental data \cite{PBM11}. 
It would therefore be not surprising that 
the role of di-neutron in the pair transfer reaction
has not yet been fully clarified. 

One such example is a relative importance of the one-step (the direct 
pair transfer) process and the two-step (the sequential pair transfer) 
process. In heavy-ion pair transfer reactions of stable nuclei, both  
processes are known to play a role \cite{EJR98}. 
For weakly-bound nuclei, most of the intermediate states for the two-step 
process are likely in the continuum spectra. 
It is still an open question how 
this fact, together with the $Q$-value matching condition,  
alters the dynamics of the pair transfer reaction of neutron-rich 
nuclei \cite{VS12}. 

On the other hand, 
the cross sections for the pair transfer reaction of the Borromean 
nuclei, $^{11}$Li and $^6$He, have been measured recently
\cite{TAB08,Oga99,Raa99,Gio05,CNS08}. 
The data for the $^1$H($^{11}$Li,$^9$Li)$^3$H reaction at 3 MeV/nucleon 
indicate that the cross sections are indeed sensitive to the pair 
correlation in the ground state of $^{11}$Li \cite{TAB08}. 
That is, the experimental cross sections can be accounted for 
only when the $s$-wave component is mixed in the ground state 
of $^{11}$Li by 30-50\%. 
Another important finding in this measurement is that significant 
cross sections were observed for the pair transfer process to 
the first excited state of $^9$Li \cite{TAB08}, 
which has made a good support for 
the idea of phonon mediated pairing mechanism \cite{PBVB10}. 

Further theoretical studies are apparently necessary in order to understand 
the connection between 
the two-neutron 
transfer process and 
the di-neutron correlation in Borromean 
nuclei. This has been left for future investigations. 

\section{Two-proton and two-neutron decays of nuclei beyond the drip lines}

In the Coulomb breakup process discussed in Sec. 2, 
the ground state wave function of a two-neutron halo 
nucleus is firstly 
perturbed by the external electromagnetic field of the 
target nucleus. It may thus not be easy to disentangle the 
di-neutron correlation 
in the ground state from that in the excited states. 
The two-proton radioactivity, that is, a spontaneous emission of 
two valence protons, of proton-rich nuclei 
\cite{PKGR12} 
is expected to provide a good tool to probe the di-proton 
correlation in the initial wave function. 
An attractive feature of this phenomenon is that the two valence protons 
are emitted directly  
from the ground state even without any external perturbation. 

Very recently, the ground state {\it two-neutron} emissions 
have also been observed, {\it e.g.,} in $^{10}$He
\cite{Golovkov09,Johansson10,Johansson10-2,Sid12,Kohley12}, 
$^{16}$Be \cite{Spyrou12}, 
$^{13}$Li \cite{Johansson10-2,Kohley13}, and
$^{26}$O \cite{Lunderberg12,Caesar13,Kondo15,Kohley13-2}. 
This is an analogous process of the two-proton radioactivity, 
corresponding to a penetration of two neutrons over a centrifugal 
barrier. Since the long range Coulomb interaction is absent, one may 
hope that the ground state correlation can be better probed by studying 
the energy and the angular correlations of the emitted neutrons, as 
compared to the two-proton decays. 

\begin{figure}
\centering
  \includegraphics[clip,width=0.4\textwidth]{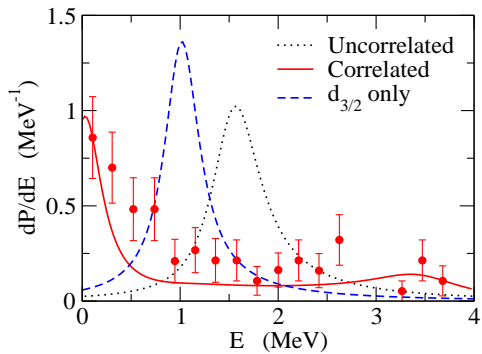}
\caption{
The decay energy spectrum for
the two-neutron emission decay of $^{26}$O, obtained with the
$^{24}$O+$n$+$n$ three-body model.
For the presentation purpose, 
a width of 0.21 MeV has been introduced to the spectrum. 
The dotted line is for the uncorrelated spectrum, while the 
solid line shows the correlated spectrum for the 0$^+$ state. 
The dashed line shows the spectrum obtained by including the pairing 
correlation only for the 
[$d_{3/2}]^2$ configuration. 
The experimental data, normalized to the unit area, are taken 
from Ref. \cite{Lunderberg12}. }
\label{fig:26o}
\end{figure}

Figure \ref{fig:26o} shows the calculated 
decay energy spectrum
of $^{26}$O obtained
with the $^{24}$O+$n$+$n$ three-body model for 
$^{26}$O \cite{SH15,HS14}. 
The calculations are carried out using the Green's function 
method as explained in Refs. \cite{SH15,EB92,HS14,HS14-2}, together 
with a density-dependent contact neutron-neutron interaction, $v$. 
In this formalism, 
the decay energy spectrum is given by, 

\begin{equation}
\frac{dP}{dE}=\sum_k|\langle\Psi_k|\Phi_{\rm ref}\rangle|^2
\,\delta(E-E_k)
=
\frac{1}{\pi}\Im \langle \Phi_{\rm ref}|G(E)|\Phi_{\rm ref}\rangle,
\label{decayspectrum}
\end{equation}
where $\Psi_k$ is a
solution of the three-body model Hamiltonian with
energy $E_k$ and $\Phi_{\rm ref}$ is the wave function
for a reference state. 
The reference state can be taken rather arbitrarily as long as it has
an appreciable overlap with the resonance states of interest.
Here, we employ 
the uncorrelated two-neutron state 
in $^{27}$F for it with the
$|[1d_{3/2}\otimes1d_{3/2}]^{(I=0)}\rangle$ configuration, which is
the dominant configuration in the initial state of the proton 
knockout reaction of $^{27}$F to produce $^{26}$O. 
In Eq. (\ref{decayspectrum}), $G$ is the correlated two-particle 
Green's function calculated as 
\begin{equation}
  G(E) =
  (1+G_0(E)v)^{-1}G_0(E)
  =G_0(E)-G_0(E)v(1+G_0(E)v)^{-1}G_0(E), 
\label{decayenergy}
\end{equation}
with the uncorrelated Green's function, $G_0$, given by 
\begin{equation}
G_0(E)=\int \hspace{-0.5cm}\sum_{\rm 1,2}\frac{|j_1j_2\rangle\langle
j_1j_2|}
{\epsilon_1+\epsilon_2-E-i\eta},
\label{green0}
\end{equation}
where $\eta$ is an infinitesimal number and 
the sum includes all independent two-particle states 
including both the bound and the continuum single-particle states. 
To this end, 
we use the Woods-Saxon potential 
for the neutron-$^{24}$O potential 
which reproduces 
the experimental single-particle energies of 
$\epsilon_{2s_{1/2}}=-4.09(13)$ MeV and 
$\epsilon_{1d_{3/2}}=770^{+20}_{-10}$ keV for $^{25}$O \cite{Hoffman08}. 
The parameters for the density-dependent zero-range pairing
interaction are determined so as to yield the
decay energy of 30 keV.

In the figure, 
we show the spectrum for the uncorrelated case by the dotted line. 
In this case, the spectrum has a peak at $E=1.54$ MeV, that
is twice the single-particle resonance energy, 0.77 MeV.
With the pairing interaction between the valence neutrons, the peak
energy is shifted towards lower energies. 
The decay spectrum obtained by including only the 
[$d_{3/2}]^2$ configurations is shown by the dashed line. 
In this case, the peak is  
shifted by $\sim$0.5 MeV from the unperturbed peak at 1.54 MeV.  
The peak is further 
shifted downwards by the configuration mixing, and gets closer 
to the threshold energy.  
This comparison implies that the pairing correlation for 
the single configuration alone is not enough to reproduce 
the empirical decay spectrum, and the di-neutron correlations between 
the two neutrons also play an essential role. 

\begin{figure}
\centering
  \includegraphics[clip,width=0.4\textwidth]{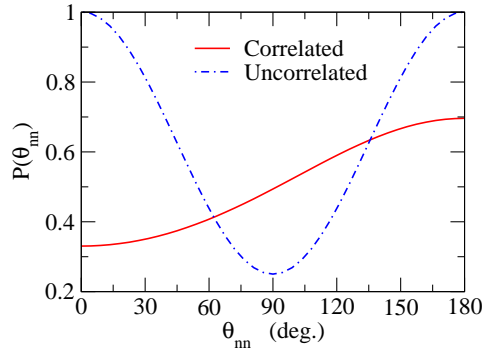}
\caption{
  The angular correlation for the two emitted neutrons from the
  ground state decay of $^{26}$O, that is, the probability 
distribution for the opening angle of the
  momentum vectors of the emitted neutrons. 
The solid and the dot-dashed lines 
denote the correlated and uncorrelated results,
respectively. }
\label{fig:26o-3}
\end{figure}

The angular distribution of the emitted neutrons can also be calculated 
with the two-particle Green's function method \cite{EB92,HS14}.  
The amplitude for emitting 
two neutrons with spin components of $s_1$ and $s_2$ and momenta 
$\vec{k}_1$ and $\vec{k}_2$ reads \cite{HS14}, 
\begin{equation}
f_{s_1s_2}(\vec{k}_1,\vec{k}_2)=
\sum_{j,l}e^{-il\pi}e^{i(\delta_1+\delta_2)}\,
M_{j,l,k_1,k_2} 
\langle [{\cal Y}_{jl}(\hat{\vec{k}}_1)
{\cal Y}_{jl}(\hat{\vec{k}}_2)]^{(00)}|\chi_{s_1}\chi_{s_2}\rangle,
\label{angularamplitude}
\end{equation}
where ${\cal Y}_{jlm}$ is the spin-spherical harmonics, 
$\chi_s$ is the spin wave function, and $\delta$ is the nuclear 
phase shift.
Here, 
$M$ is a decay amplitude calculated to a specific 
two-particle final state \cite{EB92}, 
\begin{equation}
M_{j,l,k_1,k_2}
=\langle (jj)^{(00)}|1-vG_0+vG_0vG_0-\cdots|\Psi_i\rangle 
=\langle (jj)^{(00)}|(1+vG_0)^{-1}|\Psi_i\rangle, 
\end{equation}
where the unperturbed Green's function, $G_0$, 
is evaluated at $E=e_1+e_2$.
The angular distribution is then obtained as 
\begin{equation}
P(\theta_{12})=4\pi\sum_{s_1,s_2}
\int dk_1dk_2\, |f_{s_1s_2}(k_1,\hat{\vec{k}}_1=0,k_2,
\hat{\vec{k}}_{2}=\theta_{12})|^2,
\label{angular}
\end{equation}
where we have set $z$-axis to be parallel to $\vec{k}_1$ and 
evaluated the angular distribution as a function of the 
opening angle, $\theta_{12}$,  
of the two emitted neutrons. 

Fig. \ref{fig:26o-3} shows the angular correlation so obtained.
The dot-dashed line shows 
the distribution obtained without including the 
$nn$ interaction, which 
is symmetric
around $\theta_{12}=\pi/2$.
In the presence of the $nn$ interaction, the angular distribution
turns to be 
highly asymmetric, in which the emission of two neutrons in the
opposite direction (that is, $\theta_{12}=\pi$) is
enhanced, as is shown
by the solid line.
Grigorenko {\it et al.} have also obtained a similar result \cite{Grigorenko}. 

This behavior reflects properties of the resonance wave 
function of $^{26}$O. 
That is, because of the continuum couplings, several configurations with 
opposite parity states mix coherently. 
Symbolically, let us write a two-particle wave function as,
\begin{equation}
\Psi(\vec{r},\vec{r}')=\alpha\, \Psi_{\rm ee}(\vec{r},\vec{r}')
+\beta\, \Psi_{\rm oo}(\vec{r},\vec{r}'),
\label{twowf}
\end{equation}
where $\Psi_{\rm ee}$ and $\Psi_{\rm oo}$ are two-particle wave functions 
with even and odd angular momentum states, respectively. 
The coefficients $\alpha$ and $\beta$ are such that 
the interference term in the two-particle density, 
$\alpha^*\beta\Psi^*_{\rm ee}\Psi_{\rm oo}+c.c.$, is positive 
for $\vec{r}'=\vec{r}$ while it is negative for 
$\vec{r}'=-\vec{r}$ so that  
the two-particle density is enhanced for the nearside configuration 
with $\vec{r}\sim\vec{r}'$ as compared to the far side configuration 
with $\vec{r}\sim-\vec{r}'$. 
This correlation appears in the opposite way in the momentum space. 
In the Fourier transform of $\Psi(\vec{r},\vec{r}')$, 
\begin{equation}
\widetilde{\Psi}(\vec{k},\vec{k}')=
\int d\vec{r}d\vec{r}'\,e^{i\vec{k}\cdot\vec{r}}e^{i\vec{k}'\cdot\vec{r}'}\,
\Psi(\vec{r},\vec{r}'), 
\end{equation}
there is a factor $i^l$ in the multipole decomposition of 
$e^{i\vec{k}\cdot\vec{r}}$. Since 
$\left(i^l\right)^2$ is +1 for even values of $l$ and 
$-1$ for odd values of $l$, 
this leads to \cite{SH15,HS14}
\begin{equation}
\widetilde{\Psi}(\vec{k},\vec{k}')=\alpha\, 
\widetilde{\Psi}_{\rm ee}(\vec{k},\vec{k}')
-\beta\, \widetilde{\Psi}_{\rm oo}(\vec{k},\vec{k}'). 
\end{equation}
for the two particle wave function given by Eq. (\ref{twowf}).
If one constructs a two-particle density in the momentum space with this 
wave function, the interference term therefore 
acts in the opposite way to that 
in the coordinate space. That is, the two-particle density in the 
momentum space is hindered for $\vec{k}\sim\vec{k}'$, while it is enhanced 
for $\vec{k}\sim-\vec{k}'$. 

From this argument, we can therefore conclude that, if an enhancement 
in the region 
of $\theta\sim\pi$ in the angular distribution was observed
experimentally, that would 
make a clear evidence for the di-neutron correlation in this nucleus,
although such measurement will be experimentally 
challenging \cite{Kohley15}. 

Incidentally, 
the tunneling decay of two fermionic ultracold atoms 
have been measured very recently 
\cite{Zurn13} (see also Ref. \cite{LFR15} for an 
application of the Gamow shell model to this phenomenon). 
An attractive feature of this experiment is that several parameters 
are experimentally controlable, which include the sign and the 
strength of the interaction 
between the particles and the shape of a decaying potential. 
It may be useful to carry out in future detailed analyses of 
the tunneling decay of ultracold atoms in order to shed light on 
the two-proton and two-neutron decay problems in nuclear physics. 

\section{Summary}

We have discussed possible experimental probes for the di-neutron correlation in 
neutron-rich nuclei, with which two valence nucleons are located at a 
similar position in the coordinate space. 
In particular, we have discussed the Coulomb dissociation of Borromean 
nuclei, the two-neutron transfer reactions, and the direct two-neutron decay 
of the unbound $^{26}$O nucleus. 

For the Coulomb dissociation of Borromean nuclei, even though the detailed 
distribution is difficult to extract, one can use the cluster sum rule 
(that is, the non-energy weighted sum rule) to deduce the mean value of 
the opening angle between the valence neutrons. We have demonstrated that 
the mean opening angle is 
$\langle\theta_{nn}\rangle$=74.5 $^{+11.2}_{-13.1}$ and 
65.2 $^{+11.4}_{-13.0}$ degrees 
for $^6$He and $^{11}$Li, respectively. These values are significantly 
smaller than the value for the uncorrelated distribution, that is, 
$\langle\theta_{nn}\rangle$=90 degrees, clearly indicating the existence 
of the di-neutron correlation in these Borromean nuclei. 

For the two-neutron transfer reactions, 
the sensitivity of transfer cross sections to the pairing correlation 
has been well recognized for a long time. However, the reaction dynamics 
is rather complex, and the relation to the di-neutron correlation has not yet 
been clarified completely. Given the new experimental data for the 
two-neutron transfer reactions of the Borromean nuclei, 
it is now at a good time to discuss these reactions and clarify the reaction 
dynamics with the di-neutron correlation. 

For the direct two-neutron decay, we have discussed 
the recent experimental 
data of the decay energy spectrum for the unbound $^{26}$O 
nucleus. We have shown that the decay energy spectrum can be 
accounted for only with the di-neutron correlation caused by 
a mixing of many configurations including the continuum.
We have also discussed 
the angular correlations of the emitted two neutrons. We have 
argued that the di-neutron correlation enhances an emission of 
the two neutrons in the opposite direction (that is, 
the back-to-back emission), 
and indeed our three-body model calculation has revealed such feature. 
If the enhancement of the back-to-back emission will be 
observed experimentally, 
it will thus provide a direct evidence for the di-neutron correlation. 

Even though we did not discuss them in this paper, there are other 
possible probes for the di-neutron correlation. Those include 
the nuclear breakup reaction \cite{Assie09}, the $(p,d)$ scattering 
at backward angles \cite{Horiuchi07,Suda}, and 
the knockout reactions of 
Borromean nuclei \cite{Kondo10,Kobayashi12,Aksyutina13,Uesaka}. 
It would be extremely intriguing if a clear and direct evidence 
for the di-neutron correlation could be experimentally obtained in near 
future using 
also these probes.


\end{document}